\documentclass{article}

\usepackage{arxiv}

\usepackage[utf8]{inputenc} 
\usepackage[T1]{fontenc}    
\usepackage{hyperref}       
\usepackage{url}            
\usepackage{booktabs}       
\usepackage{amsfonts}       
\usepackage{nicefrac}       
\usepackage{microtype}      
\usepackage{lipsum}		
\usepackage{graphicx}
\usepackage{natbib}
\usepackage{doi}

\usepackage{pdflscape} 

\usepackage{color}

\title{Wind-Informed Rapid Flight-Planning in Complex Urban Topologies via Machine Learning and Experimental Validation}
\renewcommand{\headeright}{ }
\renewcommand{\shorttitle}{ }

\definecolor{crassred}{RGB}{117, 13, 13}

\hypersetup{
pdftitle={A template for the arxiv style},
pdfsubject={q-bio.NC, q-bio.QM},
pdfauthor={David S.~Hippocampus, Elias D.~Striatum},
pdfkeywords={First keyword, Second keyword, More},
}

\begin{document}

\begin{center}
{\large\textsc{Wind-Informed Rapid Flight-Planning in Complex Urban Topologies\\ via Machine Learning and Experimental Validation}\\\vspace{2em}}
\hfill
Peter~I.~Renn~\textsuperscript{\textdagger1} \hfill
Alejandro~A.~Stefan-Zavala\textsuperscript{*\textdagger1} \hfill
\hfill

\hfill
Julian Humml\textsuperscript{1} \hfill
Sabera~Talukder\textsuperscript{1} \hfill
Aysha AlMazrouei\textsuperscript{2} \hfill
\hfill

\hfill
Kresna Aji\textsuperscript{2} \hfill
Debashisha Mishra \textsuperscript{2} \hfill
Emanuele Panizio \textsuperscript{2} \hfill
\hfill

\hfill
Jennifer~Simonjan \textsuperscript{2} \hfill
Yisong~Yue \textsuperscript{1} \hfill
Morteza~Gharib \textsuperscript{1} \hfill
\hfill

{\tiny
*: corresponding author: \href{mailto:aastefan@caltech.edu}{\texttt{aastefan@caltech.edu}},
\textdagger: contributed equally\\

\textsuperscript{1}: California Institute of Technology, Pasadena, CA, USA\\
\textsuperscript{2}: Technology Innovation Institute, Abu Dhabi, UAE\\

}
\end{center}

\begin{abstract}

Advanced air mobility operations hold the potential to enhance and expand regional transportation of both people and goods in populated areas. However, hazardous flight conditions arising from interactions between wind and the built environment remain a significant challenge for aerial vehicles in urban settings. This work proposes a novel framework towards safe flight planning of aerial vehicles in windy urban environments. A learning-based surrogate model is trained to rapidly predict flow fields from readily available information such as building geometry and incident wind. This surrogate prediction is used to calculate a volumetric flight challenge scalar field based on critical flow parameters and proximity to structures. A safe, flow-informed flight trajectory is then identified through a cost-minimizing pathfinder. The complete system is demonstrated experimentally through flight tests of a micro aerial vehicle through a model urban geometry placed in a large fan-array wind tunnel. Comparing this approach to trajectories generated without knowledge of the wind field, we find the flow-informed approach reduces undesired vehicle displacement and improves flight stability. This work is among the first practical demonstrations of safe, wind-aware methodologies for advanced air mobility in urban environments. 

\end{abstract}



\section{Introduction}

With applications ranging from emergency medical extractions to aerial public transportation, advanced air mobility (AAM) systems are transforming local and regional logistics and transportation \citep{vinogradov2026corridor, goyal2022advanced, raza2025advanced}. However, existing technical challenges block the widespread adoption of AAM technologies \citep{watkins2020ten}. The implication of atmospheric wind and weather conditions on flight safety remains a significant challenge for autonomous flying systems \citep{jones2022physics, mohamed2023gusts}. 
Further complicating this issue, many of the most impactful applications for AAM feature a combination of complex urban or natural topologies and high-density populations; the former can lead to accelerated wind speeds, increased turbulence levels, and unexpected gusts, while the latter demands strict safety and risk guarantees for real-world deployment. These nonlinear flow features can result in deviations from the desired flight path, posing a safety risk to potential passengers and individuals in the surrounding area. 

With better knowledge of the wind field, autonomous flying systems may be able to more safely navigate complicated fluid flows arising from complex terrain or urban topologies.  Conventional numerical approaches simulating urban wind fields through computational fluid dynamics (CFD) can provide accurate, high-fidelity, high-resolution data. \citet{geng2026urban} used numerical wind simulations in urban environments to quantify flight risk factors based on built geometries. \citet{pensado2024turbulence} used a reinforcement learning approach to simulate turbulence-aware path planning in urban environments, restricting flight in regions with numerically simulated turbulent kinetic energy above a certain threshold.  \citet{cao2026estimating} proposed a wind-informed risk-mitigating flight planning approach using simulated data, avoiding regions with high wind speeds and turbulence levels. \citet{habib2026wespr} demonstrated a flight planning approach penalizing paths through regions with high wind speed and direction misaligned with the target trajectory using CFD estimates, implementing the method using micro-UAVs and household box-fans. While invaluable in fluid mechanics research, conventional CFD simulations remain too time-intensive to be practically deployed in real-time engineering applications.

In recent years there has been significant interest in data-driven methods, such as neural networks and neural operators, as surrogate models for fluid flows in time-sensitive engineering applications \citep{brunton2020machine}. This includes approaches explicitly focused on urban environments \citep{karadag2026machine}. Surrogate models in this context typically take building geometry and a minimal representation of time-averaged wind velocity (e.g., single velocity measurement from free-stream, data from sparse sensors distributed around environment, etc.) as input, and are trained to predict the time-averaged wind field in the given environment. Previous works have featured models such as  convolutional neural networks (CNNs) \citep{wang2025evaluating}, U-Nets \citep{low2022fastflow, vargiemezis2025large}, Fourier neural operators \citep{peng2024fourier}, and generative models \citep{giral2025generative}. \citet{folk2024learning} developed a framework enabling fast wind field predictions in 2D without prior knowledge of global urban geometries, using on-board sensors to estimate surrounding buildings and local wind conditions in addition to free-stream wind conditions. However, this framework was only demonstrated using low fidelity two-dimensional flow data generated through simplified numerical methods intended exclusively for graphical applications \citep{stam2023stable}. In a following work, the authors apply a similar framework to identify safe and energy-efficient paths through simulated two-dimensional urban wind fields using a model predictive path integral optimal controller, but allow the planner to directly access the numerical data rather than surrogate model predictions \citep{folk2025flight}.

While not explicitly trained to model flows through built environments, \citet{lin2026reconstructing} trained a transformer-based surrogate on a large dataset of high-quality CFD simulations across large mountain ranges. \citet{achermann2024windseer} also demonstrated a method for predicting volumetric wind fields over complex natural terrain with the model informed by measurements taken by a small unmanned aerial vehicle (UAV), notably demonstrating this framework experimentally through field testing.

\begin{figure}
\centering
\includegraphics[width=\textwidth]{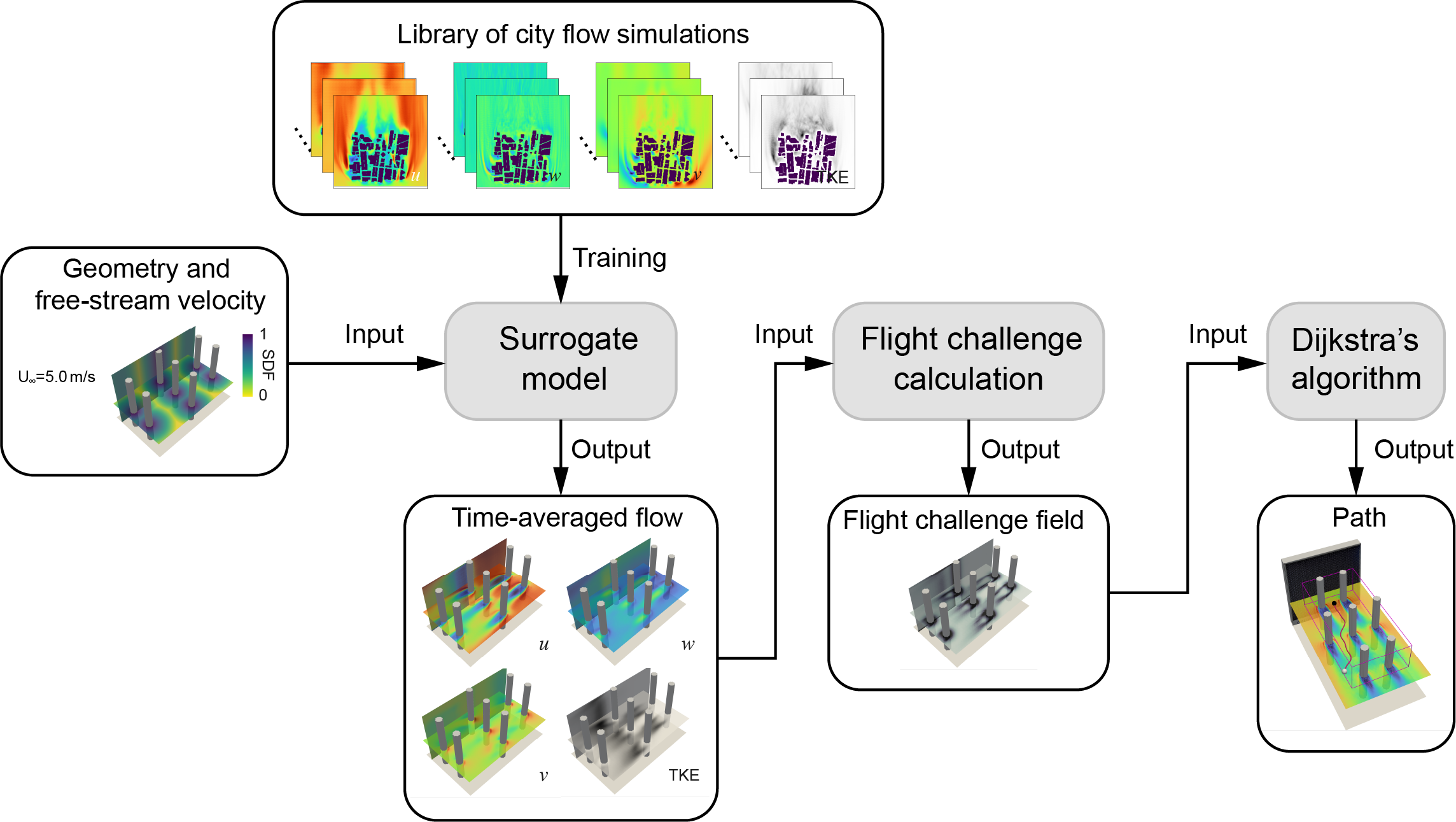}
\caption{Proposed pipeline for AAM flow-informed flight planning. A library of time-resolved flow fields are used to train a surrogate model, capable of rapidly producing accurate predictions from the known geometry and incident wind. This prediction is used to calculate a flight challenge field, with which a cost-minimizing pathfinder identifies safe flight trajectories.}
\label{fig:intro}
\end{figure}

In this work, we present a flow-informed flight planning methodology using data-driven surrogate models to identify and execute safe paths through complex urban environments in extreme wind conditions, as illustrated in Figure~\ref{fig:intro}. This real-time-capable trajectory generation is validated through experimental flight tests performed in a large-scale wind tunnel facility, establishing the first real-world demonstration of wind-aware flight planning through urban topologies. It is additionally among the first works to employ a fluid flow surrogate model in a practical, real-world setting. 

As shown in figure \ref{fig:intro}, the proposed framework uses a U-Net fluid flow surrogate model to predict volumetric time-averaged flow field quantities from the known geometry of buildings and incident wind velocity. This prediction is used to calculate a volumetric flight challenge metric, representing the potential aerodynamic hazard in various regions of the allowed flight corridor. A cost-minimizing pathfinding algorithm is then applied to the flight challenge field to identify the safest, flow-informed trajectory between any two points in the region of interest. 

First, we discuss the performance of our data-driven U-Net-based surrogate model, trained on a large library of Lattice-Boltzmann simulations over real-world urban geometries. We then evaluate the system experimentally by flying predicted paths through representative geometries in front of a large fan-array wind tunnel (FAWT). We examine the stability of trajectories generated with wind fields from both CFD and the proposed surrogate model, and compare this performance with a wind-naive alternative. Lastly, we analyze our results with respect to the simulated flow fields to examine the critical parameters relevant for flight stability.

\section{Methods}
\subsection{Flight Path Generation}

The sensitivity of aircraft to fluid disturbances across scales remains an area of active research \citep{mohamed2023gusts, jones2022physics}. In this work, a novel "flight challenge metric" is proposed to act as a basic scalar cost map describing the potential for hazardous flow conditions for aircraft throughout a given volume. This flight challenge cost map is intended to work in a time-averaged sense; rather than predicting an individual transient gust or disturbance, it aims to identify regions where gusty or otherwise adverse wind conditions may be present. To this end, the flight challenge metric was formulated through a combination of the magnitude of the velocity field gradients, the turbulent kinetic energy (TKE) field, and a Gaussian obstacle proximity mapping, as given by

\begin{equation}
\mathcal{F}(x,y,z)=
    w_1(w_2 \tilde{G} + \tilde{k})^2 + w_3D  + w_4D(w_2 \tilde{G} + \tilde{k})+ w_0.
\label{eqn:fcm}
\end{equation}

Here, $\tilde{G}$ and $\tilde{k}$ are the min-max normalized magnitude of the velocity gradient and TKE fields, respectively. $D$ is the proximity penalty, generated by convolving a Gaussian kernel with a three-dimensional binary representation of the building geometry, and the $w$ terms are weights to balance the respective contributions (with $w_0$ acting as a non-zero bias to discourage circuitous routes). 

The first term of Equation~\ref{eqn:fcm} ($w_1(w_2 \tilde{G} + \tilde{k})^2$), consisting of a squared weighted sum of the normalized magnitude of the velocity gradient and normalized TKE, aims to characterize the flight challenge given the flow physics in a specific region. Sharp velocity gradients and shear layers pose significant risk to aircraft, as vehicles must rapidly adjust to compensate for changing incident wind conditions. The turbulent kinetic energy field approximately represents the variance of velocity fluctuations in a given flow. TKE is a limited representation, as it does not characterize all quantities that determine the effect of turbulence on a specific vehicle (e.g., length scale, power spectrum). However, the TKE field still provides useful and important information in characterizing potential flight hazard due to turbulence. 

The weighted sum of the normalized TKE and magnitude of velocity gradient is squared to emphasize the more dominant and challenging flow features, and further penalize regions containing both sharp velocity gradients and high TKE (e.g., shear layers). This same squared weighted sum also appears in the third term of Equation~\ref{eqn:fcm} ($w_4D(w_2\tilde{G} + \tilde{k})$), where it is multiplied by the proximity penalty to add additional penalties to challenging flow physics in close proximity to obstacles. The cells representing obstacles in the flight challenge scalar fields were set to an arbitrarily chosen very large number to prevent paths from colliding with solid objects.

With given volumetric flight challenge scalar fields, paths were identified using existing implementations of Dijkstra's algorithms \citep{veiga2025hybrid}. This approach, performed offline, identifies minimum-cost paths between two points. Using flight challenge as cost in Dijkstra's algorithm produces flow-informed flight trajectories.

\subsection{Experimental Setup}

\begin{figure}
\centering
\includegraphics[width=0.5\textwidth]{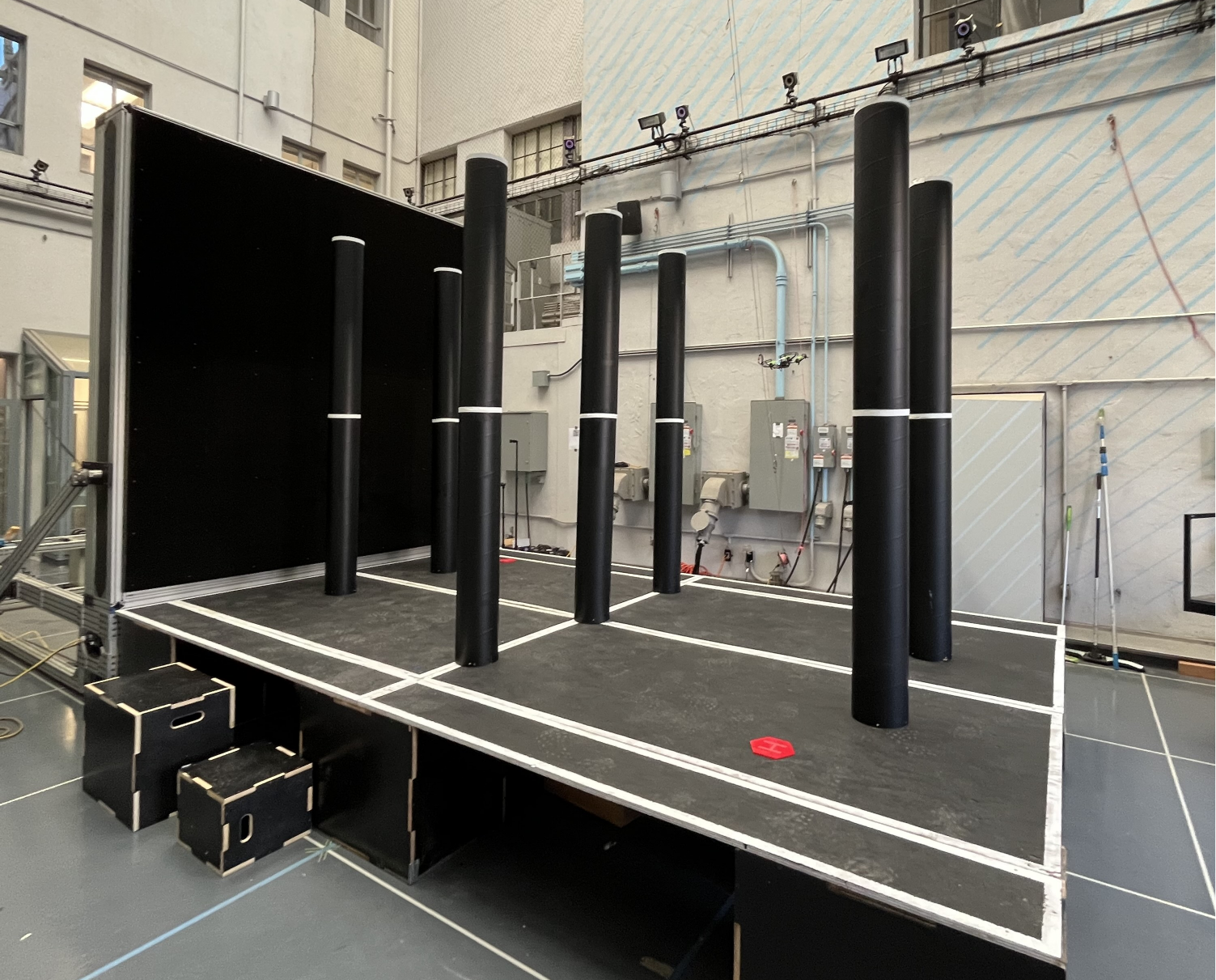}
\caption{Experimental facility used for testing.}
\label{fig:CAST}
\end{figure}

The proposed system was demonstrated experimentally in a large fan-array wind tunnel facility at the Center for Autonomous Systems and Technologies (CAST). This facility can produce wind speeds up to \textbf{15 m/s}, with a test section of 3 m $\times$ 3 m. A false floor was constructed in front of the fan-array wind tunnel, on which large obstacles were placed to represent urban geometries, as seen in Figure~\ref{fig:CAST}.

Crazyflie 2.1 Brushless quadrotors were used to fly and characterize trajectories throughout the 
building geometries and associated wind fields, as seen in-flight in Figure~\ref{fig:CAST}. Localization for these micro aerial vehicles was resolved using an array of OptiTrack motion capture cameras, with the interface running through the Crazyswarm~2 API \citep{crazyswarm}.  

In this work, we present results in the context of two distinct obstacle geometries. For each geometry, three cases are shown: surrogate-informed, CFD-informed, and flow-naive. The surrogate-informed case follows the full system pipeline as shown in Figure~\ref{fig:intro}; it uses the building geometry and known wind speed to predict the flow field, derives the volumetric flight challenge metric from this prediction, and then calculates the cost-minimizing trajectory to navigate this scalar field. The CFD-informed case uses a direct CFD simulation of the building geometry to calculate the flight challenge field and associated trajectory. Given that the surrogate model is trained on CFD data, the CFD-informed trajectory represents the best case performance of this approach, where the surrogate model prediction exactly matches its training distribution. The flow-naive case has no knowledge of the flow field, and is calculated by setting the $w_1$ and $w_4$ terms in Equation~\ref{eqn:fcm} to zero. To mitigate the impact of randomness and transient flow features, three identical trials were performed for each case.

\subsection{Surrogate Model}

\begin{figure}
\centering
\includegraphics[width=0.85\textwidth]{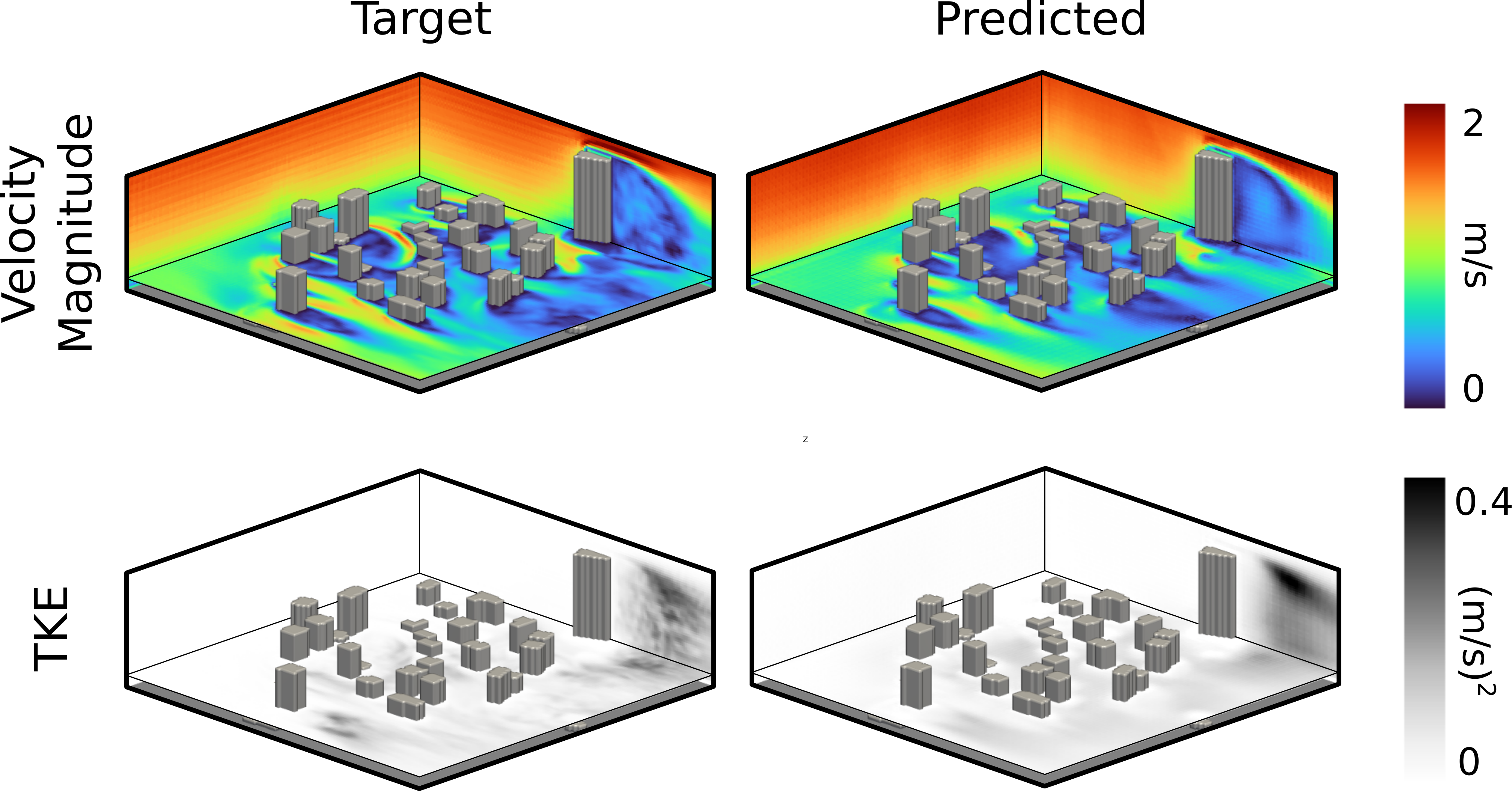}
\caption{Target and predicted fields of velocity magnitude and TKE for one sample of the holdout set. The fields are cropped from
$256\times 256\times 80$ cells to $171\times171\times50$ cells (0.684 $\times$ 0.684 $\times$ 0.2 km) for visualization.}

\label{fig:pred-vs-target}
\end{figure}

\subsubsection{Architecture and Training}
A U-Net architecture was selected to predict time-averaged flow fields, including TKE. U-Net architectures have been previously shown to effectively model fluid flows; the spatial detail-preserving characteristic skip connections aid in reconstructing fine scale flow features \citep{renn2026data}. The model input consists of a signed distance field encoding building geometries within a volume of dimensions 256 $\times$ 256 $\times$ 80. The ground plane was excluded from the distance field calculation. Feature-wise Linear Modulation (FiLM) is used to condition the model on the incident free-stream velocity, which is assumed to always flow from the $x=0$ (first volume axis) plane, with direction exactly along the positive $x$-axis. 

The model contains 9.9 $\times 10^7$ trainable parameters, and was trained on 3072 time-averaged flow fields consisting of volumetric three-component velocity and TKE. A set of 256 samples was used for validation, with an additional 64 samples held out for testing. The U-Net was set to train with a learning rate of $10^{-4}$ for 128 epochs on an NVIDIA H100 GPU, with the final model meeting the early-stopping condition after 110 epochs (approximately 15 hours of wall-clock time).

The model's performance was quantified using the normalized $\ell_2$-norm, where the error of a single 
sample is given by $
E_\mathrm{sample} = {
 ||\mathbf{u}_\mathrm{target} 
    - \mathbf{u}_\mathrm{predicted}||_2}\ /\ {||\mathbf{u}_\mathrm{target}||_2}
$ .
The final model scored a test error of 8.71\%. 
Figure~\ref{fig:pred-vs-target} shows velocity magnitude and TKE for 
one of the 64 samples in the holdout set. Velocity magnitude is used in this figure
for conciseness, while actual samples consist of all three velocity components and
TKE. Each channel and its error for the same sample are shown in Appendix~\ref{app:pred-vs-target}

\subsubsection{Data Generation}
\label{sssec:data}

A dataset of 3392 time-averaged flow fields was constructed to train, test, and validate the surrogate model. This data was generated using XLB, an open source GPU-accelerated Lattice-Boltzmann solver \cite{ataei2024xlb}. This solver was applied to 73 real-world urban geometries from around the globe, each sample spanning a square footprint of 500 m $\times$ 500 m. The geometries were extracted from the OpenStreetMap database through the Overpass API (accessed October 2025), and chosen to ensure structures would fit within the confines of the surrogate model input volume size. 

Simulations were run across geometries at random incident wind speeds (ranging between 0 and 6.5 m/s) and directions. Each simulation performed 3 $\times 10^5$ time-steps, with time-averaged quantities calculated every 2000 steps starting at step 1.5 $\times 10^5$. The velocity profile at the domain inlet was set to approximate the atmospheric boundary layer, as according to \citet{richards1993appropriate}. Due to the scale of the urban geometries and limitations with GPU memory, the simulations were formed with a uniform grid spacing of 4 m. This spacing resulted in an undesirable relaxation time approaching the traditional limit of 0.5, potentially reducing the stability and fidelity of the generated data. However, none of the simulations included in the final dataset contained the numerical errors associated with instabilities in Lattice-Boltzmann solvers, suggesting that the resulting time-averaged fields may serve as fair approximations of the flow field.

\section{Results}

\subsection{Flight Experiments}


\subsubsection{Scenario 1}
\label{sssec:scenario1}

Scenario~1 consists of two obstacles made of stacked boxes. Both obstacles are half a meter wide, and are 2.2 meters and 1.5 meters tall. The 2.2 meter tall obstacle has boxes of decreasing size stacked to produce a tiered shape, commonly seen in buildings with multiple setbacks. The obstacles are placed on an elevated platform that aligns the test-section floor with the bottom of the fan array's outlet. Incident wind was generated through the fan array at approximately 5 m/s; this inlet wind speed was also used to generate CFD and surrogate model estimates for the flow field and the resulting flight challenge metric. 

\begin{figure}
\centering
\includegraphics[width=0.5\textwidth]{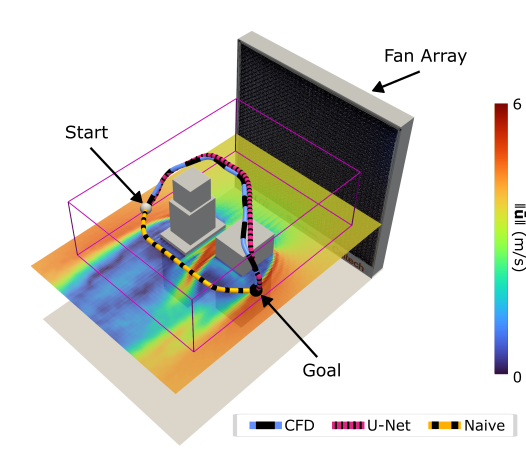}
\caption{Overview of geometry for Scenario 1 with model buildings, fan array wind tunnel, and planned paths shown. A sliced plane of the velocity magnitude is overlaid. The magenta rectangular prism represents the allowed flight corridor. }
\label{fig:gharibland_overview}
\end{figure}

Figure~\ref{fig:gharibland_overview} shows the geometry, wind field, and planned paths for Scenario 1. The path naive of the flow field is the shortest, flying immediately downstream of the obstacles. The paths informed by the proposed flight challenge metric through CFD and U-Net based wind field estimations are longer, however entirely avoid the wakes downstream of the structures. The two flow-informed paths appear very similar, as would be expected given the accuracy of the deep learning surrogate model. Given the expected presence of transient flow features, each of the desired paths shown in Figure~\ref{fig:gharibland_overview} was performed three times to better characterize performance. 

\begin{figure}
\centering
\includegraphics[width=0.5\textwidth]{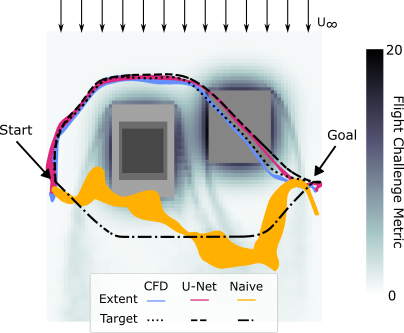}
\caption{Extent of positions in the streamwise plane across the three trials for the three paths. The desired paths are overlaid on the flight challenge metric of the bottom-most plane of the allowed flight region. The building geometries are shaded based on the respective height, to provide further context for the paths shown. Note that the CFD and U-Net informed paths pass above the building on the right-hand side.}
\label{fig:gharibland_topdown}
\end{figure}

Figure~\ref{fig:gharibland_topdown} shows a top-down view of the full extent of positions taken across all three trials, with the respective desired paths overlaid on a slice of the flight challenge metric taken at the bottom of the allowed flight corridor. It is worth emphasizing that the two-dimensional slice shown in figure \ref{fig:gharibland_topdown} does not convey the three-dimensional nature of the trajectories, especially for the two flow-informed paths (i.e., CFD and U-Net informed) which prescribe large changes in altitude to navigate around adverse flight conditions. The slice of the flight challenge volumetric field shown, being taken from the bottom plane of the flight corridor, is most relevant for the flow-naive path which experiences relatively little change in altitude. 

There exists a clear qualitative difference between the flow-informed and naive paths. The flow-informed paths (i.e., CFD and U-Net informed) appear consistent across the three trials with relatively thin traces representing the full extent of positions held. As also seen in Figure~\ref{fig:gharibland_overview}, both flow-informed cases fly upstream of the tiered building closest to the starting point while increasing in altitude, then fly directly over the building closer to the goal and reduce altitude. The trials for these paths only deviated slightly from their respective targets. 
Alternatively, Figure~\ref{fig:gharibland_topdown} shows the flow-naive paths vary significantly between trials, as represented by the thick and irregularly shaped region showing the extent, with none of the three able to closely follow the target trajectory. Notably, the regions where the naive paths show the most variance appear to correlate with the regions of higher magnitude flight challenge metric shown.

Quantifying the average magnitude of the displacement from the target in the three approaches, it is found that the CFD and surrogate model based approaches had a mean displacement of 7.7 cm and 7.4 cm respectively. The flow-naive approach achieved a mean displacement of 18.8 cm, more than twice that of flow-informed paths. Further, it should be noted that the extent shown in Figure~\ref{fig:gharibland_topdown} is only the range of positions occupied by the centroid of the vehicle rigid body; when running the flow-naive approach, the propeller guards of the vehicle were observed colliding with the tiered building in multiple trials. 

\begin{figure}
\centering
\includegraphics[width=0.5\textwidth]{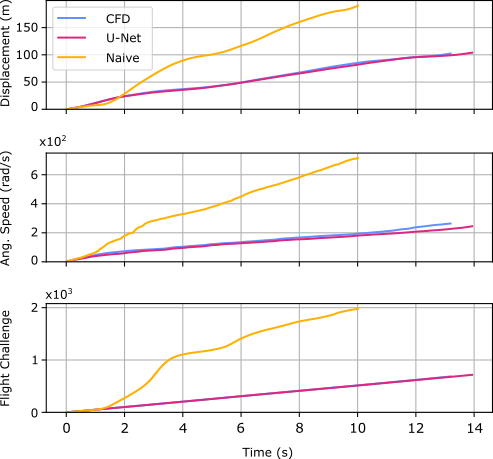}
\caption{Accumulated disturbance and flight challenge metrics as a function of flight time averaged across three trials for each of the approaches given. The magnitude of the displacements and magnitude of the angular velocity provide related but distinct characterizations of vehicle stability throughout flight.}
\label{fig:gharibland_cumulative}
\end{figure}

Figure~\ref{fig:gharibland_cumulative} shows the cumulative sum of the magnitude of the displacement, magnitude of angular velocity, and flight challenge metric averaged across the three trials for each approach. The flight challenge metric accumulation was calculated based on the CFD-generated flight challenge scalar field using the actual recorded trajectories (rather than target trajectories). While displacement is an important metric for safety and stability, as seen in Figure~\ref{fig:gharibland_topdown}, smaller disturbances can be compensated for through changes in the angular velocity. While they may not result in major deviations from the target path, these instabilities still impact overall safety and quality of the flight, and can be characterized by examining the magnitude of the angular velocity.

Although the flow-naive path is the shortest of the three shown in Figure~\ref{fig:gharibland_cumulative} by several seconds, this approach experiences significantly more displacement and angular velocity than the two flow-informed trajectories. Notably, while the CFD and U-Net paths show a relatively linear increase in deviations and flight challenge, the naive trajectory appears highly non-linear in the accumulation of all three quantities. There also appears to be a distinct correlation between the three quantities. This is most apparent examining the two inflection points which appear for all three quantities of the flow-naive case at approximately 3 seconds and 5 seconds into the flight. This relation implies that the flight challenge metric performs as desired: flying through regions with larger flight challenge values results in more deviations and higher displacement from the target path.

\subsubsection{Scenario 2}

\begin{figure}
\centering
\includegraphics[width=0.5\textwidth]{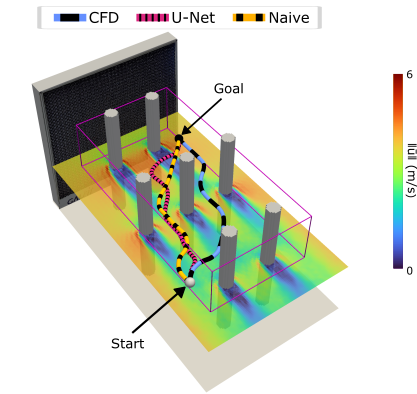}
\caption{Overview of geometry for Scenario 2. The magenta rectangular prism represents the allowed flight corridor. }
\label{fig:tubecity_overview}
\end{figure}

An offset grid of uniform circular cylinder obstacles was built in front of the wind tunnel to provide a more complicated and structured testing environment, as seen in Figure~\ref{fig:tubecity_overview}. Scenario 2 was designed to evaluate the performance of the surrogate-model based flight planning approach when complicated flow phenomena cannot be avoided entirely, as was possible in Scenario 1. To this end, the allowed flight region was constrained to the outermost edges of the cylinders as indicated by the magenta rectangular region highlighted in Figure~\ref{fig:tubecity_overview}. As in the previous scenario, these tests were performed with a constant incident wind speed of 5 m/s.

\begin{figure}
\centering
\includegraphics[width=0.8\textwidth]{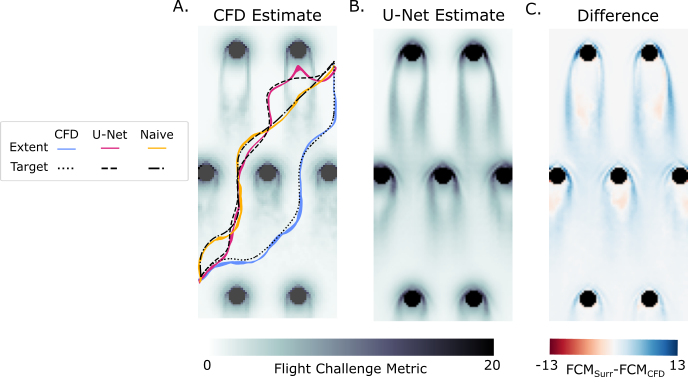}
\caption{(a.) Extent of positions in the streamwise plane across the three trials for the three paths. The desired paths are overlaid on the flight challenge metric of the bottom-most plane of the allowed flight region generated through numerical simulation. (b.) Flight challenge metric of bottom-most plane of allowed flight region generated through deep learning surrogate model (i.e., U-Net). (c.) Difference between the surrogate model and CFD-based flight challenge metric predictions. }
\label{fig:tubecity_topdown}
\end{figure}

Figure~\ref{fig:tubecity_topdown}a shows the target paths along with the range of positions taken by each respective trajectory across the three trials. Notably, the surrogate model-informed path deviates significantly from the CFD-informed path, with the surrogate model-informed trajectory passing directly through the near-wake region of the right upper-most cylinder. As would be expected, the actual flown trajectories for the surrogate-informed path deviate from the target in this near-wake region, moving very close to the upstream structure in all three trials. This is the most notable deviation across the three paths across this streamwise plane, resulting in a mean displacement of 6.7 cm for the surrogate-informed trajectory. The CFD-informed and naive paths experienced mean displacements of 5.4 cm and 5.5 cm, respectively. 

The large deviation unique to the surrogate-informed trajectory suggests a systematic issue with the surrogate model prediction, as the flight challenge metric and planning approach function as intended for the CFD-informed paths. Figure~\ref{fig:tubecity_topdown}b shows a direct comparison (relative to Figure~\ref{fig:tubecity_topdown}a) of the same two-dimensional slice of the flight challenge scalar field produced by the surrogate model, with Figure~\ref{fig:tubecity_topdown}c showing the difference between the two fields. 

While the fields follow similar geometries, the surrogate model prediction of flight challenge metric generally has higher magnitude and appears smoother than the CFD-derived field. The magnitude of the flight challenge metric in the shear layers is larger in the surrogate prediction than it is in the CFD estimate. This is unexpected, given that the surrogate-informed path routed the drone directly through the shear layer. Looking at the difference between the fields (Figure~\ref{fig:tubecity_topdown}c) along the CFD-informed path, the largest difference appears to be in the region far downstream of the uppermost right cylinder, near the second row. The flight challenge field produced through the surrogate model prediction shows a smooth gradient of non-zero values throughout this region, whereas the field derived from CFD has a more distinctly defined wake region with near-zero challenge outside. The surrogate-informed trajectory then passes directly through the near-wake recirculation bubble, and the surrounding shear layers, to avoid the downstream region containing artificially inflated flight challenge values. Overly smooth predictions are a common shortfall in deep learning models, having been previously noted in the context of surrogate models for fluid mechanics \citep{oommen2026learning, renn2026data}.

\begin{figure}
\centering
\includegraphics[width=0.5\textwidth]{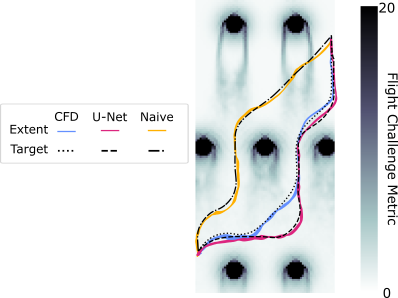}
\caption{Flight challenge metric, target paths, and range of positions with flight challenge metric recalculated with higher weight on velocity gradient.}
\label{fig:tubecity_topdown_boost}
\end{figure}

This issue can be mitigated with further tuning of the flight challenge metric parameters. The parameters used to this point were set manually using the CFD based flight challenge metric fields. By slightly adjusting the weight of the magnitude of the velocity gradient relative to TKE in the flight challenge metric formulation (increasing $w_2$ in Equation~\ref{eqn:fcm}), the surrogate-informed path follows a trajectory similar to that of the CFD-informed approach, as seen in Figure~\ref{fig:tubecity_topdown_boost}. With this adjustment, the surrogate-informed and CFD-informed trajectories have average displacements of 4.9 cm and 5.3 cm, respectively (the flow-naive path is unchanged, with average displacement of 5.5 cm). While tuning flight challenge metric parameters may solve this problem, it does not fix the underlying cause. The differences between the surrogate model and CFD flow field estimates could lead to unforeseen issues, some of which may not be solved by simply tuning existing flight challenge metric weights.

\section{Discussion}

This work presents a flow-informed approach to flight planning that mitigates risks associated with advanced air mobility operations in urban environments. This approach uses a U-Net-based surrogate model trained on CFD data to estimate time-resolved flow quantities (i.e., velocity, TKE), which are then used to calculate a derived flight challenge metric that considers velocity gradients, turbulence, and proximity to structures, as well as interactions between these terms. A basic cost-minimizing path planner is then applied to identify paths with minimal flight challenge. This system was  demonstrated experimentally in a large-scale fan-array wind tunnel facility, with model buildings and a micro aerial vehicle. 

Two scenarios with distinct geometries were considered. In each scenario, three identical trials for each of three different trajectories were executed: a surrogate-based trajectory using the full system; a reference trajectory using flow estimates from CFD instead of the surrogate model, and a baseline flow-naive approach that ignores wind and only considers proximity to obstacles. The surrogate-based full-system approach was shown to be effective in both scenarios, albeit requiring some parameter tuning to compensate for overly smooth model predictions.

While the approach presented in this work was observed to be effective, there are still limitations to address with further research. The surrogate model used in this approach produced overly smooth predictions around obstacle wakes. As seen in Scenario~2, this issue can lead to flight plans that cross high-shear regions which should have been avoided. Recent work by \citet{oommen2026learning} demonstrated adversarially trained neural operators as a means to mitigate smoothing and improve flow field predictions with little additional inference cost.

Additionally, the challenge associated with traversing a wind field is highly dependent on the geometry and capabilities of a specific vehicle. This extends beyond an aircraft's relative sensitivity to a generic flight challenge formulation. Instead,the magnitude, length scale, and direction of disturbances should be considered against the
aerodynamic profile of each aircraft in a tailored flight challenge metric.
Tailoring flight challenge metrics will require characterizing the
flight envelope and disturbance sensitivity of a given aircraft.
Alternatively, the same basic approach can be modified to simply identify "no fly" regions based on the minimum capabilities of allowed aircraft, rather than producing specific flight paths. For this, the flight challenge metric would be modified to assume worst-case aerodynamic interactions across all scales, with a threshold applied at a conservative value to identify turbulent and dangerous "no-fly" zones applicable to a wide range of aircraft. 

While there exist many potential improvements and extensions of this work, it presents an important step towards safe and resilient advanced air mobility operations. Through these experiments, we have demonstrated how data-driven physics-informed approaches to flight planning improve the safety and stability of low-altitude flight. 

\section{Data Availability}
Surrogate training data or flight logs are available upon request. 

\section*{Acknowledgments}

This work was supported by the Technology Innovation Institute (TII), UAE and the Center for Autonomous Systems and Technologies (CAST) at Caltech. ASZ thanks Skylar Wei and Matt Anderson for support with Crazyflie MAV equipment. Acknowledgement hash: 27d64cb5a58c7f2057a0ebf1979a975ce99a042aaf182a7f52c24f63859048ca.

\bibliographystyle{unsrtnat}
\bibliography{references}  

\newpage
\appendix

\begin{landscape}
\section{Holdout Set Sample}
\label{app:pred-vs-target}

\vfill
\begin{figure}[h!]
\centering
\includegraphics[width=1.3\textwidth]{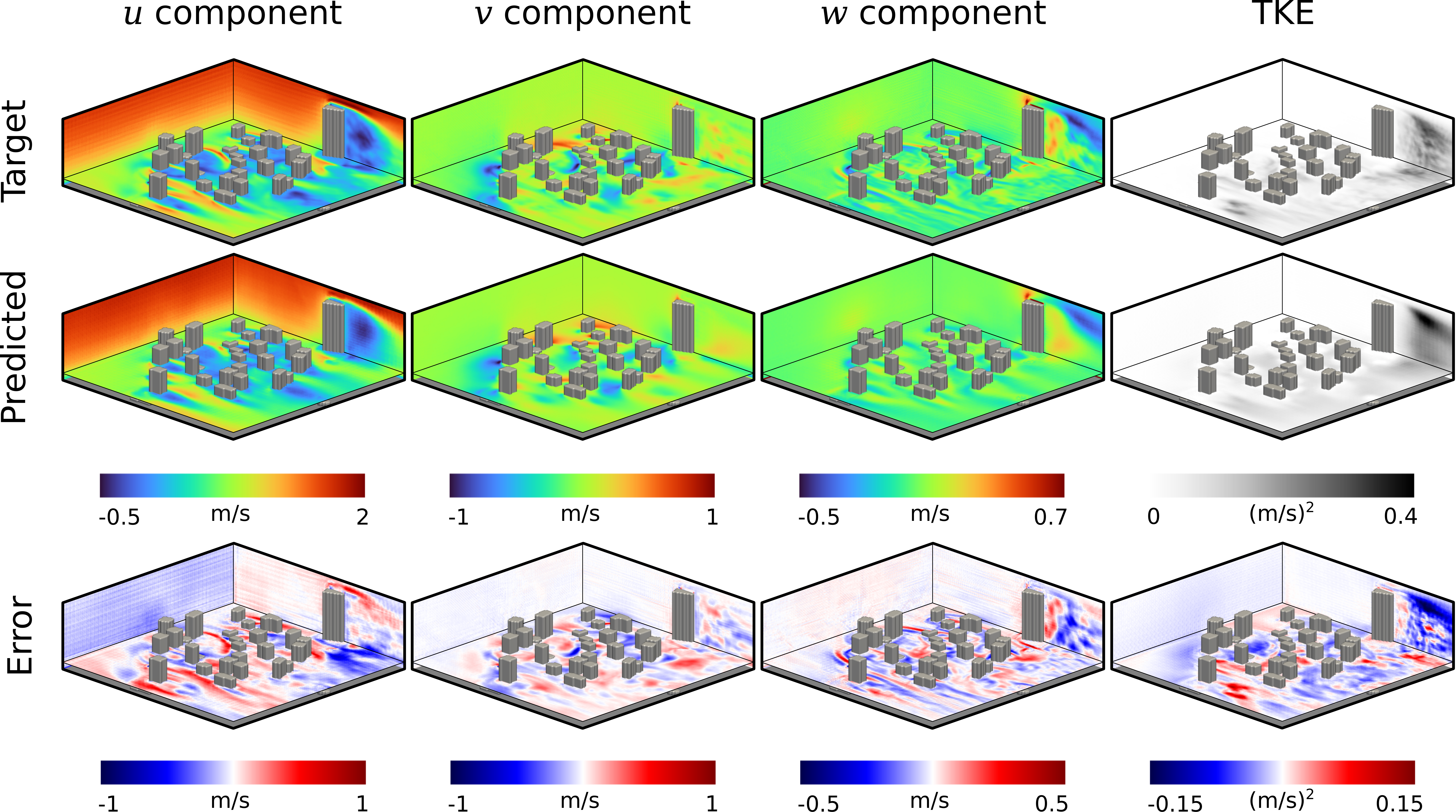}
\caption{Target, prediction and error fields of each component for one sample of the holdout set, the same sample as shown in Figure~\ref{fig:pred-vs-target}.
The fields are cropped from
$256\times 256\times 80$ cells to $171\times171\times50$ cells for visualization.}

\label{fig:pred-vs-target-full}
\end{figure}
\vfill
\end{landscape}

\end{document}